\documentstyle[aps,multicol,epsf]{revtex}

\begin{document}
\draft

\title{A Scale-free Network with Boolean Dynamics as a Function of Connectivity}

\author{A. Castro e Silva$^{1,\ast}$, J. Kamphorst Leal da Silva$^{1,\dagger}$ and
 J.F.F. Mendes$^{2,\ddagger}$}

\address{
$^{1}$ Departamento de F\'\i sica,
Universidade Federal de Minas Gerais\\
Caixa Postal 702, 30.123-970, Belo Horizonte/MG, Brazil\\
$^{2}$ Departamento de F\'\i sica, Universidade de Aveiro\\
Campus Universit\'ario de Santiago, 3810-193 Aveiro, Portugal\\}

\maketitle

\begin{abstract}
In this work we analyze scale-free networks with different power law
spectra $N(k) \sim k^{-\gamma}$ under a boolean dynamic,
where the boolean rule that each node obeys is a function of its
connectivity $k$. This is done by using only
two logical functions (AND and XOR) which are controlled by a parameter $q$.
Using damage spreading technique we show that the Hamming distance and the number
of 1's exhibit power law behavior as a function of $q$. 
The exponents appearing in the power laws
depend on the value of  $\gamma$.
\end{abstract}


\begin{multicols}{2}

\narrowtext

\section{Introduction}

This paper blends two promising concepts of complex systems, namely the 
boolean networks\cite{kau0,kau1}  and  models of growing networks\cite{ba99,baj99}. 
The former was introduced by Kauffman \cite{kau0,kau1} in 1969. One
important thing about Kauffman's model is that it was very well received
not just by the physical but by the biological community as well. The 
simplicity and richness of behaviors of this model created a huge field of research,
with many types of different boolean networks \cite{carlos}. 
Moreover, this model became a good
candidate to explain biological problems such as cell differentiation, gene expression,
protein interaction and genetic regulatory networks \cite{amt,wei,kada}.
 A boolean network is a complex dynamical system constituted of logical variables connected
by logical functions. The simplest boolean model has two parameters, the number of
logical variables $N$ and the number of inputs of their boolean functions $K$, where
$K$ varies from $K=1$ (the function has just one input) to $K=N$
(all variables connected with all the others including itself). The boolean functions
are chosen randomly in the beginning and they are kept during the dynamics.
  The network updates
synchronously, meaning tha all nodes refresh their state at the
same time. This kind of network exhibit an ordered dynamic for $K=1$ and a chaotic one when
$K=N$. However, the network self-organizes showing an interesting level of order and
complexity at $K=2$, the edge of chaos. In addition to Darwin's
natural selection and random mutation, Kauffman's idea is that self-organization and 
random dynamic can be
responsible for the complexity observed in nature.

The models of growing networks introduced by Barab\'{a}si and Albert \cite{ba99,baj99}
are in the other side. In these works was proposed a very elegant way of creating
a self-organized scale-free structure. In contrast to most types of networks previously known
, the scale-free ones exhibit a power-law distribution of connections, $P(k) \sim k^{-\gamma}$
 where $\gamma$ is the so called scale-free exponent \cite{dm03}. This behavior is
similar to what happens with lengths in fractal structures. The mechanism that makes
possible the arising of such property is the so called {\it preferential attachment},
that is, the probability of an old node to receive a new link from a new node
is proportional to the number of pre-existent connections in this node. The growing
nature of these networks allows the study of a wide type of networks, such as
the World Wide Web, neural, social relations, disease spreading, voting, citations of
scientific papers, movie's actors interconnections, etc
\cite{atb,ws98,ls98,red98,hppj98,ajb99,hub,baam99,new99,bw00,asbs00,dmj00}. The second
characteristic of a scale-free network is the {\it small world effect}, which means that 
the shortest distance between two nodes is of the order of $\ln (N)$, where $N$ is the
size of the network. In other words,
no matter how large the network could be, any two nodes are connected via a
relatively small chain of links. For instance in WWW any document can
be reached with less of 19 mouse clicks. The third feature of scale-free graphs 
is the very good tolerance to random removal of a significant fraction of its nodes 
conjugated with a high vulnerability to directional attacks to the most connected nodes (hubs). 

Although these two branches of research are in plenty expansion, with a huge
amount of publications, the combination of both are just in the beginning, and
little is known about the behavior of the random networks under
scale-free topology.

Since scale-free networks describe more realistically the
interactions among members of any types of network, the boolean dynamic can help
to understand how some sort of dynamics flow inside those groups
with that topology. From the biologic point of view, the introduction of
the scale-free structure can be a decisive step to describe in a
qualitative and quantitative fashion what is observed in genetic networks, metabolic 
pathways and protein interaction networks.
All these networks have developed from very simple ones under the preassure 
of natural selection.   
They have evolved by gradual changes (mutations) that, simultaneously 
keep its functionality. Therefore we expect that scale-free networks are a more realistic 
description since they are growing graphs. It is worth 
mentioning, that scale-free network are more robust to random errors than homogeneous random 
models. Experimental evidences have been found in cellular networks of some living beings, 
as example the yeast {\it Saccharomyces cerevisiae} \cite{yeast}.

Few papers have been published using boolean dynamics in scale-free networks. Fox and Hill 
used a random dynamics in a network with maximum connectivity $K_{max}=30$ to simulate 
the regulation of gene expression\cite{fox}. Aldana and Cluzel analytically demonstrated 
the existence of a phase transition for values of the scale-free exponent
in the open interval $(2,2.5)$ was analytically demonstrated in the random
dynamics \cite{aldana}.

In this work we study  scale-free networks with a deterministic boolean dynamics
using numerical simulation.
We consider that the dynamics  is driven only by AND and XOR functions. These
functions are controlled by an external parameter $q$. This simple dynamics allows us
to simulate large networks. We consider the Hamming distance $D$ and the number of 1's
$M$ for asymptotic times and different values of $q$. After averaging for several initial
conditions we find that these quantities vary with $q$ as power laws.

In the following section we define
the scale free networks and  the boolean dynamics.  Section III present our results for
 the numerical simulations, which are then discussed in Section IV.  
 A brief summary is given in Section V.

\section{The networks and the dynamics}

The computer simulation was performed in two major parts. The
first part was the growth of the network, which corresponds to the static part of the simulation
since the network is unchangeable during the dynamics.
We grew  scale-free networks with minimum connectivity $k_{min}=1$ and
$k_{min}=2$ by
using the {\it Growing Network with Re-direction} algorithm\cite{gnr}.
The minimum connectivity corresponds to the smallest number of links
that a node can have. The two kinds of network are grown as follows:
\begin{itemize}
\item $k_{min}=1$ -- a new node is linked to only one old node; we select
an old node with uniform probability; then the link with the old node is established with probability $1-r$ or
it is redirected
to the ancestor of the old node with probability $r$;
\item $k_{min}=2$ -- a new node is linked to two old nodes; we  select
an old node with uniform probability; then one link with the old node is established with probability $1-r$ or
it is redirected
to one of the two ancestors of that node with probability $r$; we repeat the same procedure for
the other link.
\end{itemize}
 In both cases, the initial condition consists of three nodes
with cyclic connections.
 This algorithm creates a scale-free network with
$\gamma=r^{-1}+1$. For instance, when $r=0.5$, the network has $\gamma=3$,
corresponding to a growth with linear preferential attachment.
 We use different values of $r$ for the two kinds of
 networks.

The second part of the simulation is the dynamic itself. Once the network is made of $N$ nodes connected by links,
to each node $i$ is assigned a logical variable $\sigma_i(t)$.
 The state of the network at time $t$ is represented by a set of boolean variables
$(\sigma_1(t), \sigma_2(t), \sigma_3(t), ..., \sigma_N(t))$. At each time step, the state of a node $i$ is
defined in the following way:

i) for $k_{min}=2$, $\sigma_i(t+1)$ is given
by $F_i(\sigma_{i_1}(t), \sigma_{i_2}(t), ..., \sigma_{i_{k_i}}(t))$. In other words, the state of
a node $i$ at $t+1$ is a function of the states at $t$ of all nodes linked with $i$, namely
$\sigma_{i_1}(t), \sigma_{i_2}(t), ..., \sigma_{i_{k_i}}(t)$, where $k_i$ is the connectivity of $i$-th node.

ii) for $k_{min}=1$, the state of a node is defined in a different way,
since there are nodes with just one link. Then $\sigma_i(t+1)$ is  given by
$F_i(\sigma_i(t), \sigma_{i_1}(t), \sigma_{i_2}(t), ..., \sigma_{i_{k_i}}(t))$. Now, the state of
a node $i$ at $t+1$ is a function of the state at $t$ of all nodes  linked to it and of its own state as well.
 In this way we have always  a function with at least two inputs.

Finally we define the function $F_i(\sigma)$:

\begin{equation}
F_i = \cases{ AND \,&if\, $k_i<q$ \cr
                XOR \,&otherwise\cr}
\label{1}
\end{equation}

where $q$ is a threshold parameter that controls how the logical functions  AND and
 XOR spread into the network.
The AND and XOR logical functions were introduced in order to simplify the
model, avoiding the necessity of defining $2^{2^k}$ different boolean relations for
each node. It is
know from previous works that the AND function leads to an ordered regime (with two
fixed points, where all variables are 0's or 1's) while the XOR introduces a more
chaotic component to the dynamic. 
These kinds of functions are necessary to study biologic networks since they are the
boolean counterparts of real reactions in cell regulatory system\cite{kau1,wei}.
Once in the scale-free network, we have many nodes
poorly connected (small $k$) and few nodes highly connected (large $k$), equation
(\ref{1}) we can set a balance between chaotic and ordered dynamics inside the
network.

An initial state $\Sigma=(\sigma_1(0),
\sigma_2(0), \sigma_3(0), ..., \sigma_N(0))$ is created by assigning randomly 0's and 1's to all nodes.
A copy
$\overline{\Sigma}=(\overline{\sigma_1}(0), \overline{\sigma_2}(0), \overline{\sigma_3}(0), ...,
\overline{\sigma_N}(0))$
of the initial state is also created and we introduce a damage  by changing the value
of one randomly chosen node.
Both the initial state and the damaged state evolve under the control of equation (\ref{1}).
 Once the new state of all nodes is calculated the entire network is
updated (synchronous update) and the system goes to the next Monte Carlo time step (mcs).
Note that, except for the random choice of the initial state,
 the dynamics is deterministic. This means that the boolean functions act with a probability
$p=1$, and this is another simplification to the dynamics. Random variables are important in order to simulate
real genetic networks under influence of many uncertainties like
biologic variability, experimental noise and interacting variables impossible
to quantify. However, this simplification allows us to consider large networks without
loss of complexity in the dynamical behavior.

\section{The simulation data and results}

 The dynamical behavior is characterized by  two quantities. The first is the
average density of 1's, namely
\begin{equation}
M(q,t)=\lim_{N\to\infty} \langle\frac{1}{N}\sum_{i=1}^N\sigma_i(t)\rangle~~.
\label{eq_M}
\end{equation}

 The second is the average of the Hamming distance, that is defined by
\begin{equation}
D(q,t)=\lim_{N\to\infty} \langle\frac{1}{N}\sum_{i=1}^N|\sigma_i(t)-\overline{\sigma_i(t)}|\rangle~~.
\end{equation}

 Here, $\langle\ldots\rangle$ is the average for different initial conditions.
 An initial condition is given not only by the initial states ($\Sigma$, $\overline{\Sigma}$) but also
 by  one set of links of a grown network with a specific $\gamma$.

\begin{figure}
\vspace{0.2cm}
\epsfysize=9cm
\epsfxsize=7.5cm
\centerline{\epsfbox{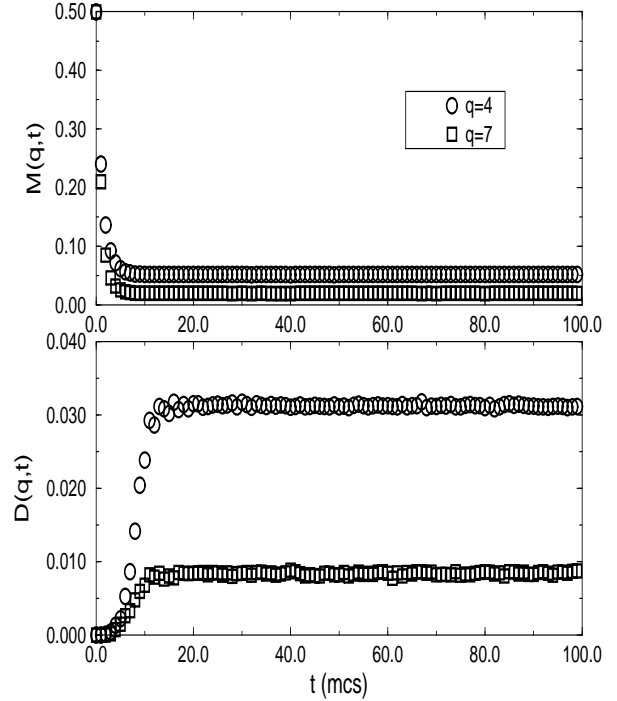}}
\vspace{0.2cm}
\caption{Plots of $M(q,t)~vs~t$ and $D(q,t)~vs~t$ for a network with $N=8\times
10^4$, $k_{min}=1$ and $r=0.5$.  }
\label{fig1}
\end{figure}

After a very short transient (less than $30$ time steps) these quantities reach the stationary values $M(q)$ and $D(q)$,
as it is shown in Fig.~\ref{fig1} for a network with $N=8\times 10^4$ nodes, $k_{min}=1$ and $r=0.5$. The stationary
values, defined as 

\begin{eqnarray*}
M(q)&=& \lim_{T \to \infty} 1/T \int_t^{t+T} M(q,t^\prime) dt^\prime~~,\\
D(q)&=& \lim_{T \to \infty} 1/T \int_t^{t+T} D(q,t^\prime) dt^\prime~~,
\end{eqnarray*}
were determined by discarding the first $50$ time steps and by making a time average until $t=200$ time steps.
 We can also see in the figure that the stationary values $D(q)$ and $M(q)$ depend on $q$.
  We are interested in this  dependence. It is worth mentioning that a similar behavior is found for networks with
  different values of $r$ and also for all networks with $k_{min}=2$.

In order to consider the finite size effects, we grew networks with $N=1\times 10^4$,
$N=2\times 10^4$,  $N=4\times 10^4$ and $N=8\times 10^4$ for networks with $k_{min}=1$
and $k_{min}=2$.  The  sample averages were performed at least with $10^2$ samples
(small $q$ and large $N$). For large $q$ we have used up to $3\times 10^4$ random
initial conditions.

\begin{figure}
\vspace{0.2cm}
\epsfysize=9cm
\epsfxsize=7.5cm
\centerline{\epsfbox{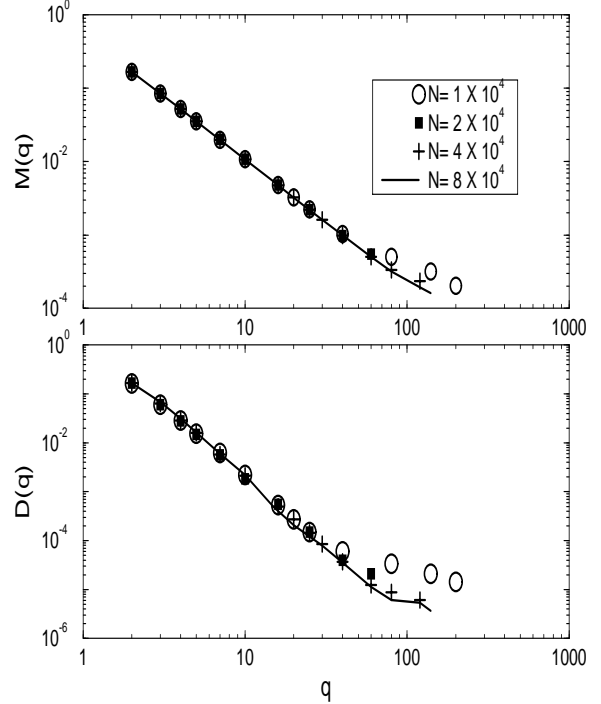}}
\vspace{0.2cm}
\caption{Log-log plots of $M(q)~vs~q$ and $D(q)~vs~q$ for a network with $k_{min}=1$, $r=0.5$ and different $N$.}
\label{fig2}
\end{figure}

We can see from Fig.~\ref{fig2} that $D(q)$ and $M(q)$ have the following asymptotic power-law behaviors:
\begin{eqnarray}
M(q)&\sim& q^{-m}\label{eqm}~~,\\
D(q)&\sim& q^{-d}\label{eqd}~~.
\end{eqnarray}

\begin{figure}
\vspace{0.2cm}
\epsfysize=9cm
\epsfxsize=7.5cm
\centerline{\epsfbox{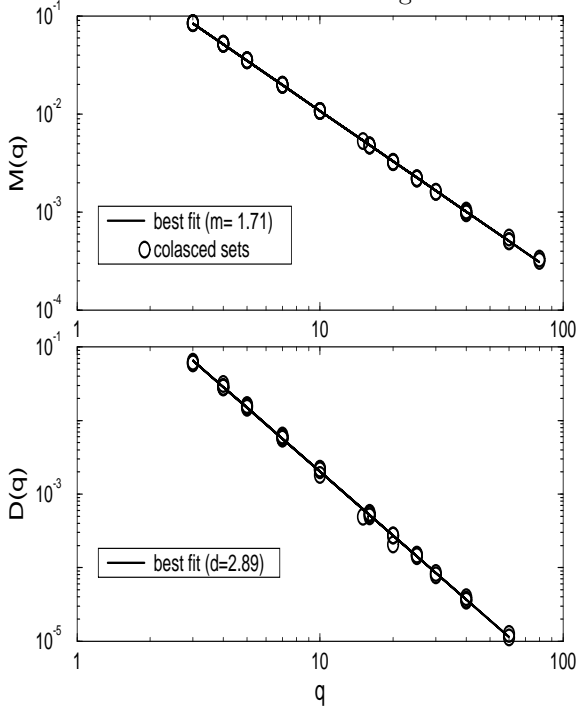}}
\vspace{0.2cm}
\caption{Log-log plots of $M(q)~vs~q$ and $D(q)~vs~q$ for a network defined in Fig.~\ref{fig2}. It shows 
the best fit of the coalesced sets.}
\label{fig3}
\end{figure}

Moreover, we can also see in the same figure, finite size effects for large $q$ by comparing the behavior of
the smallest network ($N=1\times 10^4$) with the largest one ( $N=8\times 10^4$). In order to evaluate the
exponents, we eliminate the points affected by finite size effect and coalesce all different sets. Finally
we do a best fit. This is shown in Fig.~\ref{fig3}, for the case defined in Fig.~\ref{fig2}. There, the first
point ($q=2$) was not considered in the best fit. Note that we have evaluated the exponents by considering
 approximately 2 orders of magnitude in the $q$ variable and that the fit is very good. In fact, in all fitted
 data, we obtained a correlation coefficient larger that $0.999$.

\begin{figure}
\vspace{0.2cm}
\epsfysize=9cm
\epsfxsize=7.5cm
\centerline{\epsfbox{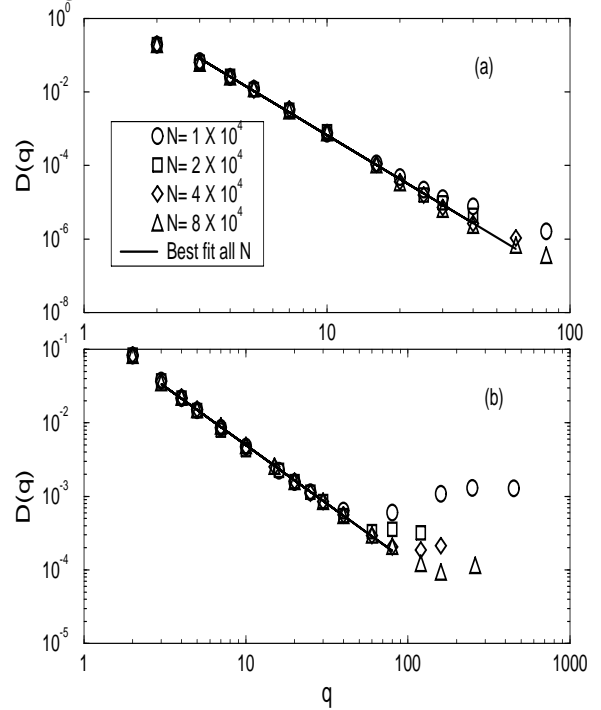}}
\vspace{0.2cm}
\caption{Log-log plots of  $D(q)~vs~q$ for a network with $k_{min}=1$ and (a)$r=0.35$ and (b)$r=0.80$.
The best fits of the coalesced sets, after the elimination of the points affected by finite size and  the discard of
the first point, furnish (a)$d=3.98$ and (b)$d=1.59$.}
\label{fig4}
\end{figure}

 Plots of $D(q)$ versus $q$ are shown in Fig.~\ref{fig4} for  networks with $k_{min}=1$ and different values of
 $r$ ((a)$r=0.35$ and (b)$r=0.80$). Note that the finite size effects are
 present for $q \approx 260$  for the
 $r=0.8$ and $N=8\times 10^4$ case. This implies that finite size effects  appear 
 for $q<<k^*$, where $k^*$ is the static cut-off. In fact, when we are evaluating the 
   connectivity distribution $P(k)$ for this case, we expect that finite size effects appear when 
   $k\sim k^*=N^r\approx 8\times 10^3$
   \cite{mendes-web}. In a numerical simulation of the exponent  $\gamma$ for
   this case, we found that the finite size effects appear for $k\sim 1000$. Therefore, 
   if we want to evaluate the exponents with 3 orders of magnitude, we must grow
   networks much larger than the ones we consider here. 

We follow this procedure to calculate all the other exponents. They are displayed in 
table~\ref{tab1}. 
 The table also show the values of the scaling-free exponent evaluated 
 numerically $\gamma_n$ and the exact value $\gamma=1+r^{-1}$. In the numerically 
 evaluation of $\gamma_n$ we considered the same networks used in the dynamics with
 $10^5$ samples. We can 
see that the exponents $m$ and $d$ change with $r$ and $k_{min}$.
It is worth mentioning that the asymptotic value of $D(q)$ is independent of 
the initial amount of damage. Only the short time behavior depends on it. In 
particular if the initial damage is larger than $D(q)$, $D(q,t)$ shows a decay 
to the stationary value.

\begin{table}
\caption{\sf Exponents $m$ and $d$ for $k_{min}=1$ and $k_{min}=2$ for
different values of $r$. It is also shown the scaling-free exponent 
evaluated numerically $\gamma_n$ and the exact  one $\gamma$.}
\label{tab1}
\begin{center}
\begin{tabular}{|c|c|c|c|c|c|c|}
 &\multispan2\hfil $k_{min}=1$\hfil \vrule&\multispan2 \hfil$k_{min}=2$\hfil\vrule& \\
\hline
 $r$    &    $m$         &    $d$      &    $m$         &      $d$    & $\gamma_n$ &$\gamma$\\
\hline\hline
$0.35$  &   2.25 (3)     &  3.98 (5)   &   2.77 (1)     &   4.10 (2)  & 3.50     &  3.86   \\
$0.5$   &   1.71 (1)     &  2.89 (2)   &   2.00 (1)     &   2.80 (3)  & 2.92     &  3.00    \\
$0.65$  &   1.38 (2)     &  2.15 (5)   &     -          &   -         & 2.52     &  2.54    \\
$0.8$   &   1.16 (3)     &  1.59 (1)   &   0.53 (1)     &   0.70 (1)  & 2.27     &  2.25    \\

\end{tabular}
\end{center}
\end{table}

\begin{figure}
\vspace{0.2cm}
\epsfysize=9cm
\epsfxsize=7.5cm
\centerline{\epsfbox{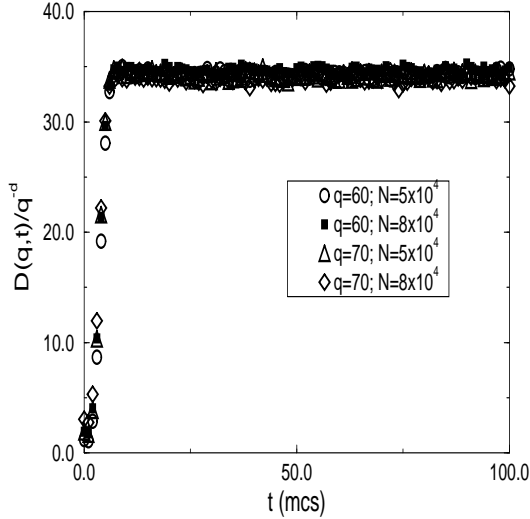}}
\vspace{0.2cm}
\caption{Temporal evolution of $D(q,t)$ for different $q$ and $N$, where
 $k_{min}=2$ and $r=0.5$. (see legend).
The $y$ axis was rescaled by $D(q,t)/q^{-d}$ in order to show the collapse.}
\label{fig5}
\end{figure}

In Fig.\ref{fig5} we show the temporal evolution of $D(q,t)/q^{-d}$, for the 
network with $k_{min}=2$ and $r=0.5$. Note that after an initial transient, all the curves 
collapse.

\section{Discussion of Results}

 In order to discuss the simulation results, we present a
 simple approximation for the evaluation of the exponents $m$ and $d$ for
 the case $k_{min}=1$,
  in which we neglect all the correlations between
 nodes. Note that argument is also valid for $k_{min}=2$. 
 The dynamics defines two sublattices of nodes depending on the
 connectivity $k$ of each node. The sublattices $a$ and $b$
 consist of all nodes with  $k<q$ and $k\geq q$, respectively. Then the
 nodes of subllatice $a$($b$) evolve by AND (XOR) operations. We are interested
 in the steady state, where the distribution of connectivity is described
 by $P(k)\sim k^{-\gamma}$. In this regime, we consider that a typical node 
 $i$ of the sublattice $a$ with connectivity $k$ has 
 $\sigma_i=0$, because the AND operation of the nodes $\sigma_i,\sigma_1\ldots\sigma_k$
 is $0$ if at least  one of these $k+1$ nodes is zero. On the other hand, a node of the 
 sublattice $b$ has a probability $1/2$ of being nonzero, because half of all possible configurations
of $\sigma_i,\sigma_1\ldots\sigma_k$, under XOR operation, give $\sigma_i=1$. The average
density can be written in terms of the sublattices as  $M(q)=M_a(q)+M_b(q)$. Since $M_a(q)\approx 0$,
we have
\begin{equation}
M(q)\approx M_b(q)=\frac{1}{2}\int_q^\infty P(k)~dk\sim q^{-(\gamma-1)}~~,
\end{equation}
implying that $m=\gamma -1$. To evaluate $D(q)$ we must consider a lattice and
its copy that have evolved from two different initial states.  Again we have that $D(q)=D_a(q)+D_b(q)$, 
where $D_j(q)$ is the contribution of the sublattice $j$($j=a,b$).
 So, we need the number of nodes that have different values
in the  lattice and its copy for each one of the sublattices.  Since the nodes of the sublattice $a$ 
 have the  value $0$ in both lattices, $D_a(q)\approx 0$. Only the nodes of the sublattice $b$ can be different 
 from 0.
 The probability that the same node be $1$ at one lattice and $0$ in its copy is $1/4$, if we
 neglect all correlations. Therefore we can write that
 \begin{equation}
D(q)\approx D_b(q)\approx \frac{1}{2}\int_q^\infty P(k)~dk\sim q^{-(\gamma-1)}~~.
\end{equation}
 Using the definition of the exponent $d$, we find that $d=m=\gamma -1$.
 From Table~\ref{tab1}, we see that the values of the exponents are different from the ones predicted
 by our approximated evaluation. This suggests that correlations are important in this problem.
 To obtain a better understanding we studied the steady-state behavior of the sublattices for  
 networks with $N=10^4$ nodes. The results of the sublattice $a$ are shown in Table~\ref{tab2} in
 percentages of the quantities characterizing the complete network.
 
  Let us first analyze the case with $k_{min}=1$. From Table~\ref{tab2}, we note that the 
  sublattice $a$
  is irrelevant for the Hamming distance independently of the value of $r$.

\begin{figure}
\vspace{0.2cm}
\epsfysize=9cm
\epsfxsize=7.5cm
\centerline{\epsfbox{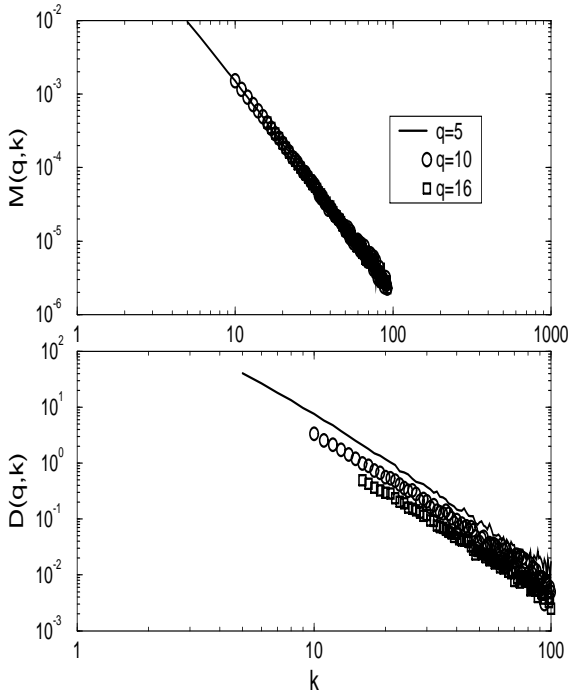}}
\vspace{0.2cm}
\caption{Log-lo plots of the sublattice $b$ quantities $M_b(q,k)~vs~k$ and $D_b(q,k)~vs~k$
 for different $q$, where
 $k_{min}=1$, $r=0.5$ and $N=10^4$.
}
\label{fig6}
\end{figure}

\begin{table}
\caption{\sf Sublattice percents of $M_a(q)$ and $D_a(q)$  for $q=5, 10$ and 16 with 
$k_{min}=1$, $N=10^4$ and different values of $r$.}
\label{tab2}
\begin{center}
\begin{tabular}{|c|c|c|c|c|c|c|}
 &\multispan3\hfil $M_a$  \hfil \vrule&\multispan3\hfil $D_a$ \hfil\vrule \\
\hline
 $q$ / r    &0.35    & $0.5$          & $0.8$     &0.35    &  $0.5$     & $0.8$    \\
\hline\hline
$5$        & 7\%    &   6\%          &  4\%        &0.03\%    &   0.03\%      &   0.01\%        \\
$10$       & 18\%   &   16\%         &  9\%        &1\%       &   0.3\%       &   0.01\%       \\
$16$       & 25\%   &   24\%         &  15\%       &4\%       &   1.3\%       &   0.3\%        \\

\end{tabular}
\end{center}
\end{table}
 On the other hand, the contribution of this sublattice for the average density is relevant. This
 contribution for $r=0.35$ is larger than the one for $r=0.8$. 
 This indicates that the sublattice $a$ is almost irrelevant for $r=1$. To get a deep insight we analized
 $M_b(q,k)$ and $D_b(q,k)$, respectively, the average density and the average Hamming distance
 of nodes with connectivity $k$ for a fixed parameter $q$. Note that both definitions involve only
  the nodes of sublattice $b$. From the numerical simulations we obtain that the behavior of these
  quantities can be described by
\begin{eqnarray}
M_b(q,k)&\sim& A_qk^{-\alpha}\label{eqmk}~~,\\
D_b(q,k)&\sim& B_qk^{-\beta}\label{eqdk}~~,
\end{eqnarray}
with $\alpha$ and $\beta$ having  values approximately equal to $\gamma_n$,
 the numerical estimate of $\gamma$. This implies that $\alpha=\beta=\gamma$.  Therefore, the sublattice
 quantities can be expressed as
\begin{eqnarray}
M_b(q)&=\int_q^\infty M(q,k)~dk&\sim A_q q^{-(\alpha -1)}\label{eqmq}~~,\\
D_b(q)&=\int_q^\infty D(q,k)~dk&\sim B_q q^{-(\beta -1)}\label{eqdq}~~.
\end{eqnarray}
 In Fig. \ref{fig6} it is shown the graph $M_b(q,k)\times k$ for  $k_{min}=1$, $r=0.5$, $N=10^4$ 
and $q=5,10,16$. 
A best fit furnishes $\alpha=2.89=\gamma_n$ and an intercept $A_q$, both independent of $q$. 
Similar results are obtained for the other values of the parameter $r$. 
 If we neglect $M_a$, then we can write that $M(q)\sim M_b$ and  $m=\alpha-1=\gamma -1$.
 Then, the exponent $m$ 
should have a value near $\gamma -1$.  This can be verified in Table~\ref{tab1} and the difference 
 between $\gamma -1$ and $m$, 
  which is around $20\%$ for $r=0.35$ and $7\%$ for $r=0.8$, is due mainly to the 
  contribution of the sublattice $a$. Since this contribution becomes smaller and the
  exponent $m$ approaches $\gamma -1 $ as $r\to 1$, we conjecture that $m=\gamma -1=1$ when
  $r=1$.
  
 Fig. \ref{fig6} displays $D_b(q,k)\times k$.
 A best fit furnishes $\beta=2.89\approx \gamma_n$ independent of $q$ and an intercept $B_q$ that depends
  on $q$. This dependence is due to the correlation between the nodes of the two sublattices.
 Similar results are obtained for different values of $r$. 
 Since $D_a\approx 0$ independently of $r$,  we have that $D(q)\approx D_b(q)\sim B_qq^{\beta -1}$, with
   $\beta-1=\gamma -1$. Then the exponent $d$ 
should have a value different from $\gamma -1$ because of $B_q$. This is confirmed in the simulations, 
 with the intercept presenting a power law behavior $B_q\sim q^{-\beta_1}$. For $r=0.5$, 
 $\beta_1\approx 1$ and we have that $d\approx\gamma$.
 When $r$ becomes larger, the straight lines
 for different $q$ in a log-log plot become very close, implying that $\beta_1\to 0$.
 Moreover, the exponent $d$ approaches $\gamma -1 $ as $r\to 1$. Then, we conjecture that
 $d=\gamma -1=1$ for $r=1$.

It is worth mentioning that for $r=1$, a new node is always redirected to the ancestor node. It means
that we have three hubs, the initial nodes, with all other nodes connected to them. In fact we
have a largest hub, with $61\%$ of the all nodes connected to them,  a smallest one ($11\%$ of
connections) and a third hub with $28\%$ of all links. During the dynamics, the XOR operation is
applied to the hubs  and  AND  to the nodes with a single link.

\begin{figure}
\vspace{0.2cm}
\epsfysize=9cm
\epsfxsize=7.5cm
\centerline{\epsfbox{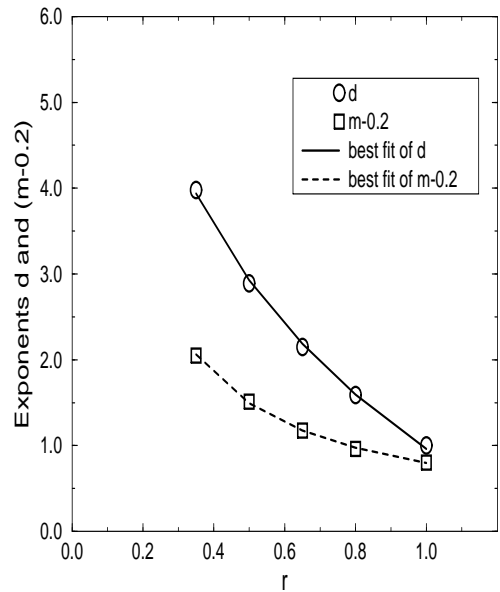}}
\caption{Exponents $d$ and $m-0.2$ as a function of $r$ for $k_{min}=1$. 
}
\label{fig7}
\end{figure}

The exponents $d$ and $m-0.2$ as a function of $r$ for $k_{min}=1$ are shown in Fig.~\ref{fig7}. Note
that we included the values $m=d=1$ for $r=1$ in the data. Although we have only 5 points, we can do a best fit.
We obtain that $d = 0.963 - 2.836 \ln (r)$ and $m-0.2= 0.795 r^{-0.908}$ with correlation
coefficients around $-0.9995$.

  Now, let us discuss the case $k_{min}=2$. We can see from Table~\ref{tab3} that the sublattice $a$ is
  now relevant. In fact, as long as $r\to 1$, this sublattice is more important than subllattice $b$. Moreover,
   the nodes with connectivity $k=2$ are responsible for the major contribution of  $D(q)$ or $M(q)$. For example,
   when we have that $r=0.8$ and $q=10$, they correspond to $88\%$ of $D$ and to $83\%$ of $M$. The scenario
   for $r=1$ is as follows: sublattice $b$ has no direct contribution ($D_b\approx 0$ and $M_b\approx 0$) and
   only the nodes with $k=k_{min}=2$ are responsible for $D(q)\not= 0$ and $M(q)\not= 0$. 
 Now, we have three hubs, the initial nodes, with all other nodes connected to them by two links.
 The largest hub and the smallest one have, respectively, $52\%$  and $17\%$ of the all nodes connected 
to them. During the dynamics, the XOR operation is applied to the hubs  and  AND  to the other nodes. 
 These arguments imply that the exponents $d$ and $m$ are not related to $\gamma$ in the same way as the 
 previous case ($k_{min}=1$).  

\begin{table}
\caption{\sf Sublattice percents of $M_a(q)$ and $D_a(q)$  for $q=10$ and $20$ or $25$  with 
$k_{min}=2$, $N=10^4$ and different values of $r$.}
\label{tab3}
\begin{center}
\begin{tabular}{|c|c|c|c|c|c|c|}
 &\multispan3\hfil $M_a$  \hfil \vrule&\multispan3\hfil $D_a$ \hfil\vrule \\
\hline
 $q$ / r    &0.35    & $0.5$          & $0.8$      & 0.35     &  $0.5$     & $0.8$    \\
\hline\hline
$10$       & 24\%    &   43\%          &  84\%     & 32\%     &   53\%      &   89\%        \\
$20$       & 17\%    &   43\%          &  -        & 22\%     &   53\%      &    -       \\
$25$       &  -      &    -            &  92\%     &  -       &    -        &   95\%        \\

\end{tabular}
\end{center}
\end{table}

\section{Summary}
We have studied a deterministic boolean dynamics with two boolean functions (AND and XOR) 
controlled by an external parameter $q$. We have considered two distinct networks with 
minimum connectivity given by $k_{min}=1$ and $k_{min}=2$. In the first case, the state 
of a node at time $t+1$ is a function of all connected nodes plus its own value at previous time $t$.
In the second case, the state of a node at $t+1$ depends only on the states of all connected 
nodes at time $t$. We have grown networks with different scale-free exponents by changing 
a parameter $r$. The finite size effect were take into account by considering networks with 
different sizes. We have shown that the density of the Hamming distance $D$ and the density of 1's 
$M$ as a function of $q$ have a power law tail for asymptotic times.

It seems that the exponents $d$ and $m$ 
characterizing the behaviors of $D$ and $M$, 
respectively,  depend on $k_{min}$ and $r$, this means that the exponents depend on the 
details of the dynamics.

ACS thanks the hospitality of the Departamento de Fisica da Universidade 
de Aveiro, where part of work took place.\\ 
We thank to G. J. M. Garcia for the critical reading of the manuscript.\\
JFFM was partially supported by the projects POCTI/FAT/46241/2002 and POCTI/MAT/46176/2002.\\
JKLS thanks to Funda\c c\~ao de Amparo \`a Pesquisa do Estado de Minas Gerais (FAPEMIG) and
to Conselho Nacional de Pesquisa (CNPq), Brazilian agencies.\\
$^{\ast}$      Electronic address: alcides@fisica.ufmg.br\\
$^{\dagger}$      Electronic address: jaff@fisica.ufmg.br\\
$^{\ddagger}$   Electronic address: jfmendes@fis.ua.pt\\

\end{multicols}

\begin{references}

\bibitem{kau0}  S. Kauffman, J. Theoret. Biol. {\bf 22}, 437  (1969).

\bibitem{kau1}  S. Kauffman {\it At Home in the Universe}
Oxford University Press, New York, 1995.

\bibitem{ba99}  A-L. Barab\'{a}si and R. Albert, Science {\bf 286}, 509
(1999). 

\bibitem{baj99} A-L. Barab\'{a}si, R. Albert and H. Jeong, Physica A {\bf 272}, 173
(1999).

\bibitem{carlos} C. Gershenson, J. Broekaert and D. Aerts, {\it Proceedings Lectures Notes in Artificial
Intelligence} {\bf 2801}, 615 (2003).

\bibitem{amt}  A. Bhattacharjya and S. Liang, Phys. Rev. Lett. {\bf 77}, 1644 (1996).

\bibitem{wei}  I. Shmulevich, E. R. Dougherty and W. Zhang, IEEE, {\bf 90}, 1778 (2002).

\bibitem{kada}  S. N. Coppersmith, L. P. Kadanoff, Z. Zhang, Phys D, {\bf 157}, 54
(2001)

\bibitem{dm03} S.N. Dorogovtsev and J.F.F. Mendes, {\it Evolution of
Networks: from biological nets to the Internet and WWW}, Oxford University Press, 2003.

\bibitem{atb}  A. T. Bernardes, D. Stauffer and J. Kertesz, Eur. Phys J. B {\bf 25},
123 (2002).

\bibitem{ws98}  D.J. Watts and S.H. Strogatz, Nature {\bf 393}, 440 (1998).

\bibitem{ls98}  J. Lahererre and D. Sornette, Eur. Phys. J. B {\bf 2}, 525
(1998).

\bibitem{red98}  S. Redner, Eur. Phys. J. B {\bf 4}, 131 (1998).

\bibitem{hppj98}  B.A. Huberman, P.L.T. Pirolli, J.E. Pitkow and R.J.Lukose, Science {\bf 280}, 95 (1998).

\bibitem{ajb99}  R. Albert, H. Jeong and A-L. Barab\'{a}si, Nature {\bf 401}, 130 (1999).

\bibitem{hub} B.A. Huberman and L.A. Adamic, Nature {\bf 401}, 131 (1999).

\bibitem{baam99} J.B.M. Barth\'{e}l\'{e}my and L.A.N. Amaral, Phys. Rev. Lett. {\bf 82}, 3180 (1999); erratum: Phys. Rev. Lett. {\bf 82}, 5180 (1999).

\bibitem{new99} M.E.J. Newman and D.J. Watts, Phys. Lett. A. {\bf 263}, 341 (1999).

\bibitem{bw00}  A. Barrat and M. Weigt, Eur. Phys. J. B {\bf 13} 547 (2000).

\bibitem{asbs00}  L.A.N. Amaral, A. Scala, M. Barthelemy, and H.E. Stanley, cond-mat/0001458.

\bibitem{dmj00} S.N. Dorogovtsev and J.F.F. Mendes, Europhys. Lett., {\bf 50}, 1 (2000).

\bibitem{fox} J. J. Fox and C. C. Hill, Chaos {\bf 11}, 809 (2001).

\bibitem{yeast} P. Uetz, Nature {\bf 403}, 623-627 (2000).

\bibitem{aldana} M. Aldana and P. Cluzel, PNAS {\bf 100}, 8713 (2003).

\bibitem{gnr}  P.L. Krapivsky and S. Redner, J. Phys. {\bf A 35}, 9517 (2002).

\bibitem{mendes-web} S.N. Dorogovtsev, J.F.F. Mendes and A. N. Samukin, Phys. Rev. E {\bf 63}
 062101 (2001). 
 
\bibitem{ashish} A. Bhan, D. J. Galas and T. G. Dewey, Bioinformatics, {\bf 18},
1486 (2002).

\bibitem{hamm} R. W. Hamming {\it Coding and Information Theory} 
Prentice Hall, NJ, 1986.


\end{references}
\end{document}